\documentclass[twocolumn]{openjournal} 

\usepackage{orcidlink} 
\usepackage[T1]{fontenc}
\usepackage{ae,aecompl}
\usepackage[T1]{fontenc}
\usepackage{ae,aecompl}
\usepackage{hyperref}
\usepackage{url}

\usepackage{longtable}
\usepackage{booktabs}
\usepackage{tabu}
\usepackage{graphicx}
\usepackage{multirow}
\usepackage{ulem}
\usepackage{color}
\usepackage{amsmath}

\def \cm{~\rm{cm}}
\def \s{~\rm{s}}
\def \km{~\rm{km}}

\def \K{~\rm{K}}

\def \erg{~\rm{erg}}

\def \yr{~\rm{yr}}

\def \kpc{~\rm{kpc}}

\usepackage{xcolor}
\definecolor{redak}{rgb}{0.9,0.15,0.05}

\shorttitle{The jets of CCSNR Cir X-1}
\shortauthors{Soker}

\graphicspath{{./}{figures/}}

\begin{document}

\title{The explosion jets of the core-collapse supernova remnant Circinus X-1}

\author{Noam Soker\,\orcidlink{0000-0003-0375-8987}} 
\affiliation{Department of Physics, Technion - Israel Institute of Technology, Haifa, 3200003, Israel; soker@physics.technion.ac.il; }

\author{Muhammad Akashi\,\orcidlink{0000-0001-7233-6871}}
\affiliation{Kinneret College on the Sea of Galilee, Samakh 15132, Israel}
\affiliation{Department of Physics, Technion - Israel Institute of Technology, Haifa, 3200003, Israel; akashi@technion.ac.il}

\date{\today}

\begin{abstract}
We propose that the recently analyzed opposite rings in the Circinus X-1 (Cir X-1) core collapse supernova (CCSN) remnant resulted from a pair of opposite jets at the final phases of the jet-driven explosion process of the progenitor of Cir X-1. We point out the similarity of the rings in the Cir X-1 CCSN remnant to a ring in the Cygnus Loop CCSN remnant. While the X-ray binary system Cir X-1 actively launches jets, no such activity exists in the Cygnus Loop. In both CCSN remnants, we attribute the rings to jets associated with the explosion process, within the framework of the jittering-jets explosion mechanism (JJEM). We also identify such a ring in the CCSN remnant 107.7-5.1, which we also attribute to exploding jets. We conduct three-dimensional hydrodynamical simulations of late jets inside an exploding massive stellar core, and demonstrate the feasibility of this scenario for ring formation. The Cir X-1 CCSN remnant has a large blowout, similar to that of the Cygnus Loop and to a large protrusion in the CCSN remnant G0.9+0.1. Based on these similarities, we suggest that other exploding jets inflated the blowout of the Cir X-1 nebula, consistent with an earlier claim regarding the formation of the blowout of the Cygnus Loop. We identify a point-symmetric structure in the Cir X-1 CCSN remnant, strengthening the JJEM. This study further demonstrates that the JJEM is a successful explosion mechanism to analyze CCSNe and CCSN remnants. 
\end{abstract}
   
\keywords{supernovae: general -- stars: jets -- ISM: supernova remnants -- stars: massive}

\section{Introduction} 
\label{sec:intro}

There is no consensus on the explosion mechanism of core-collapse supernovae (CCSNe), with two alternative mechanisms studied in more than 10 papers in 2025 alone. 

One is the neutrino-driven explosion mechanism (delayed neutrino mechanism; e.g., \citealt{Bambaetal2025CasA, Bocciolietal2025, EggenbergerAndersenetal2025, Huangetal2025, Imashevaetal2025, Laplaceetal2025, Maltsevetal2025, Maunderetal2025, Morietal2025, Mulleretal2025, Nakamuraetal2025, SykesMuller2025, Janka2025, ParadisoCoughlin2025, Vinketal2025, WangBurrows2025}). The magnetorotational explosion mechanism, which involves one pair of jets along a fixed axis that explodes the star, requires a rapidly rotating pre-collapse core  (e.g., \citealt{Shibagakietal2024, ZhaMullerPowell2024, Shibataetal2025} and references to earlier papers therein), and occurs in rare CCSNe. It asserts that the neutrino-driven mechanism is responsible for the majority of CCSNe explosions; accordingly, we grouped it with the neutrino-driven mechanism concerning the primary explosion mechanism of CCSNe.  
Some studies in the frame of the neutrino-driven mechanism attribute the extra energy of some superluminous CCSNe to a magnetar, i.e., a rapidly rotating magnetized neutron star (NS) remnant. 
However, many of these cases require explosion energies of $E_{\rm exp} \gtrsim 3 \times 10^{51} \erg$ (e.g., \citealt{Orellanaetal2025, Aguilaretal2025}, for some most recent examples), more than the neutrino-driven mechanism can supply, implying explosion by jets (e.g., \citealt{SokerGilkis2017, Kumar2025}); most magnetar models overlook this conclusion.

The other intensively studied CCSN explosion mechanism is the jittering jets explosion mechanism (JJEM; e.g., \citealt{BearSoker2025, SokerShishkin2025Vela, SokerShiran2025, ShiranSoker2026}), which has benefited from the recent identification of point-symmetric morphologies in more than 10 CCSN remnants (CCSNRs; e.g., \citealt{Soker2026G11}).

The increasing number of CCSNRs exhibiting jet-induced structural shaping has two significant implications. The first is that it relates the shaping of CCSNRs to jet-shaping of some other astrophysical objects, particularly active galactic nucleus jets that shape the interstellar and intracluster medium (e.g., \citealt{Soker2024CF, SokerShishkinW49B} for some comparisons), and jet shaping in planetary nebulae (e.g., \citealt{Soker2024PNSN, Bearetal2025Puppis}). The second implication is that jet-shaped morphologies, particularly point-symmetric morphologies, can inform our understanding of the explosion mechanism of CCSNe. Point-symmetric morphologies are those that have two or more pairs of opposite structural features that do not share the same axis through the center of the nebula or remnant; the two structural features in a pair are opposite to the center. The opposite structural features might include dense clumps; dense elongated structures termed filaments; bubbles, which are faint structures enclosed by a brighter rim; lobes, which are bubbles with partial rims; and ears, which are protrusions from the main CCSNR shell with decreasing cross-sections away from the center. 
The two implications are related, as the morphological similarity to other astrophysical objects known to be powered and shaped by jets is used to argue that jets also power CCSNe (e.g., \citealt{Soker2025G0901, Soker2024Keyhole} and references therein) within the framework of the JJEM. 
 
The main task of the JJEM studies over the past two years has been to identify CCSNRs with point-symmetric morphologies and other signatures of jet shaping. Sixteen CCSNRs were identified with point-symmetric morphologies, most of them in 2024-2025 (e.g., \citealt{Soker2025Learning, Soker2025N132D, Shishkinetal2025S147, Soker2026G11} for a list of these CCSNRs). 
The JJEM attributes points symmetrical morphologies to pairs of jets that participated in the explosion mechanism, as jittering pairs of jets (i.e., not along the same axis) shape point-symmetrical morphologies of bubbles, lobes, clumps, filaments, and ears, as three-dimensional simulations show \citep{Braudoetal2025}. 
For this study, it is particularly relevant to note that in most cases, the jets in the JJEM are not relativistic (these jets are not the jets of gamma ray bursts that are relativistic, e.g., \citealt{Izzoetal2019, AbdikamalovBeniamini2025}; \citealt{Guettaetal2020} claim that most CCSNe do not have relativistic jets).

The basic properties of the JJEM are as follows (For a review of all JJEM properties and outcomes, see \citealt{Soker2025Learning}). 
Stochastically intermittent accretion disks around the newly born NS launch $N_{\rm 2j} \simeq 5-30$ pairs of jittering jets during the explosion process. The variations in the axes of pairs of jittering jets may be fully or partially stochastic; partially stochastic behavior occurs when the pre-collapse core is rapidly rotating, so that the jittering is around this angular-momentum axis. The sources of stochastic angular momentum fluctuations are vortices in the convection zones of the collapsing core; instabilities above the NS amplify these fluctuations (e.g., \citealt{GilkisSoker2014, ShishkinSoker2023, WangShishkinSoker2024, WangShishkinSoker2025}; for such instabilities see, e.g., \citealt{Abdikamalovetal2016, KazeroniAbdikamalov2020, Buelletetal2023}).

\cite{Bearetal2025Puppis} who analyzed SNR Puppis A and \cite{Shishkinetal2025S147} who analyzed SNR S147 conclude that in some cases there are one, two, or three very powerful pairs of jets, and that one jet might be much more powerful than the opposite jet in a pair. By momentum conservation, this pair imparts a kick velocity to the NS in the direction of the weak jet; this mechanism is termed kick-BEAP for kick by an early asymmetrical pair of jets. 

 One of the challenges of the JJEM is that present numerical simulations of collapsing cores and NS formation do not obtain accretion disks. A possible reason is that present numerical simulations should include magnetic fields and do not yet have the required resolution. The JJEM requires intermittent accretion disks to launch the exploding jets. To that end, accretion disks must amplify magnetic fields (a dynamo) that rapidly reconnect to release their energy. \cite{Soker2025Learning} estimates the width of magnetic field reconnection zones to be $D_{\rm rec} \approx 0.005r \simeq 0.1 \km$ near the surface of the NS, therefore, requiring numerical resolutions several times smaller than the resolution of present CCSN simulations. \cite{Soker2025Learning} concluded that existing simulations of the CCSN explosion mechanism are still far from resolving the relevant processes of the JJEM. The JJEM must overcome this challenge.   

In a recent study, \cite{Gasealahweetal2025} detected two opposite protrusions from the nebula around the X-ray binary Circinus X-1 (hereafter Cir X-1) and termed this CCSNR the Africa Nebula, as its morphology resembles the African continent. They termed the protrusions `bubbles'; to be consistent with many earlier studies of CCSNRs, we will refer to these protrusions as `ears' (i.e., a bubble is a faint zone surrounded by a brighter rim). \cite{Gasealahweetal2025} conducted two-dimensional hydrodynamical numerical simulations and argue that jets, which the Cir X-1 X-ray binary launched after the CCSN explosion, shaped these protrusions. In this study, we compare the Africa Nebula to other CCSNRs (Section \ref{sec:CirX1}) and raise the possibility that the two jets that formed the pairs of ears were part of the explosion process (Section \ref{sec:jets}). We cannot and do not rule out the explanation of \cite{Gasealahweetal2025}, but only point to a possible alternative one. We also suggest an additional jet axis in the JJEM frame. In Section \ref{sec:Simulations} we present preliminary results from simulations we are working on. In Section \ref{sec:kick}, we discuss the kick-BEAP concerning the kick velocity of the binary system Cir X-1. We summarize in Section \ref{sec:Summary}.   

\section{Cir X-1 and other CCSNRs}
\label{sec:CirX1}

The thorough and enlightening study of the Cir X-1 CCSNR by \cite{Gasealahweetal2025}, who termed it the Africa Nebula, motivated us to examine this CCSNR and three other CCSNRs. These three CCSNRs exhibit morphological features suggesting two or more pairs of jets that exploded and shaped them. We first clarify the terminology used to describe the morphological features of the Africa Nebula.

\begin{figure*}[]
	\begin{center}
\includegraphics[trim=0.0cm 19.5cm 0.0cm 0.0cm ,clip, scale=1.00]{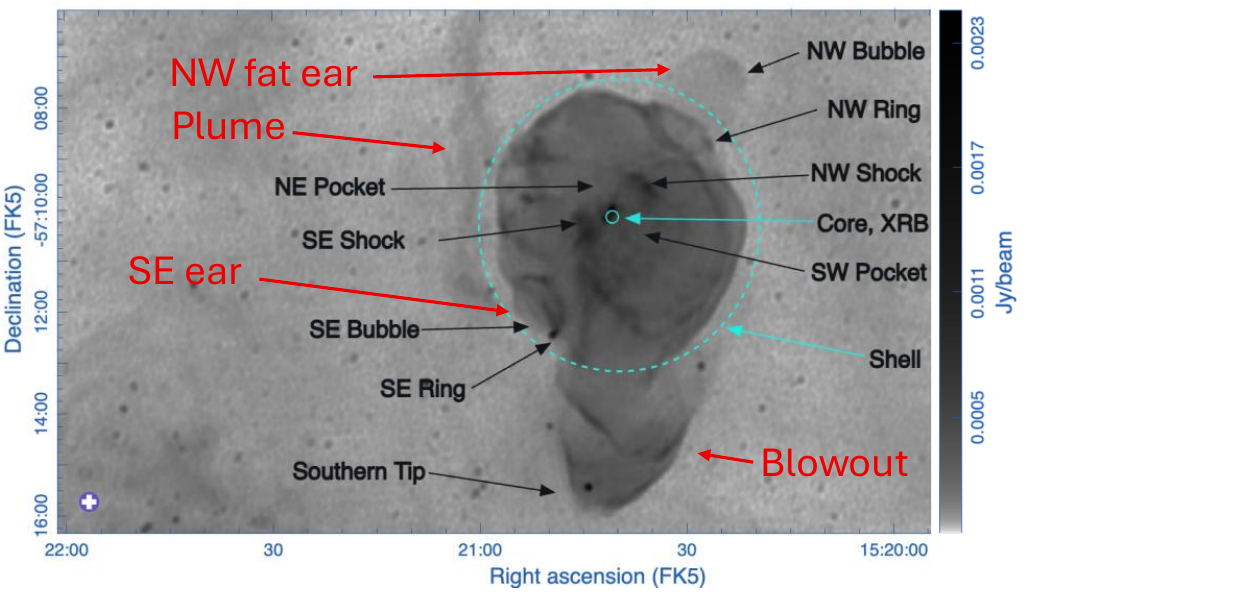} 
\caption{A figure adapted from \cite{Gasealahweetal2025}, showing a radio  MeerKAT L band image of the Cir X-1 nebula, termed the Africa Nebula. Black labeling is from \cite{Gasealahweetal2025} and red from this study; we term their bubbles ears, and define the plume and the blowout.  
}
\label{Fig:CirC1}
\end{center}
\end{figure*}

\begin{figure*}[]
	\begin{center}
\includegraphics[trim=0.0cm 5.4cm 4.2cm 0.0cm ,clip, scale=0.86]{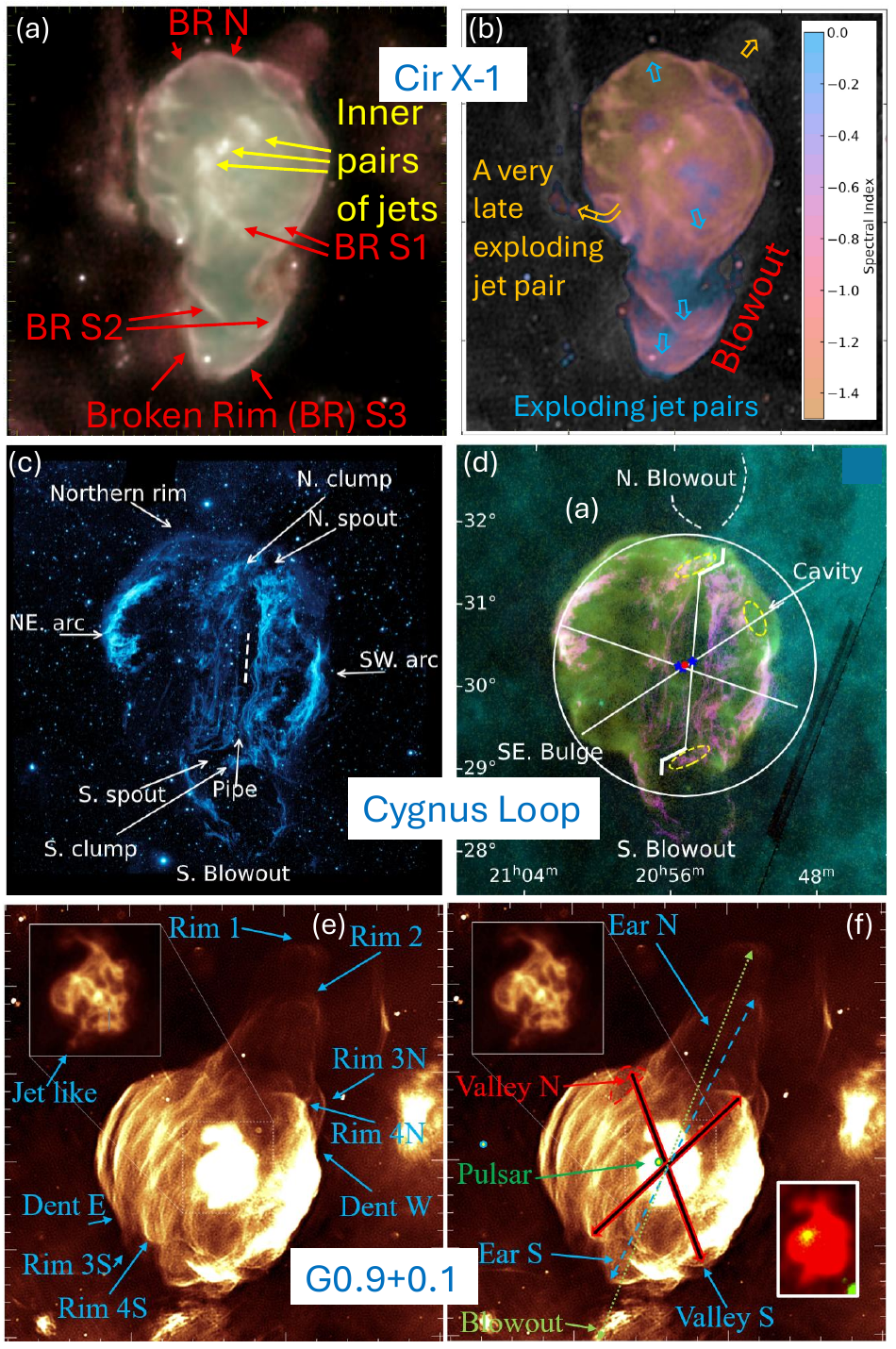} 
\caption{Comparing the Africa Nebula to two other CCSNRs. (a+b) Images of the Africa Nebula adapted from \cite{Gasealahweetal2025}. We added all marks: in yellow, features discussed by \cite{Gasealahweetal2025}; and in red, orange, and pale blue, our new definitions. (a) Three sub-bands of the radio continuum observations by colors. We identify four broken rims (BR). (b) A map combining spectral index (by color according to the color bar) and intensity (by brightness). Double-lined arrows schematically represent our suggestion for energetic exploding jets. 
(c+d) Images of CCSNR Cygnus Loop adapted from \cite{ShishkinKayeSoker2024}, with all marks from \cite{ShishkinKayeSoker2024}. (c) GALEX UV image in 175 - 280 nm (\citealt{GALEX2007}; Credit: NASA/JPL-Caltech). (d) Magenta represents visible, green represents log scale X-ray, and teal represents the AKARI $90 \mu m$ image.  
(e+f) A MeerKat radio image at 1.28 GHz of SNR~G0.9+0.1 adapted from \cite{MeerKAT2022} with all marks from \cite{Soker2025G0901}. (e) The inset on the upper left is a desaturated image of the pulsar wind nebula. (f) The inset is an image from \cite{Camiloetal2009}, which reported the discovery of the pulsar indicated by the yellow dot. \cite{Soker2025G0901} marked by the dashed pale blue and the green-dotted two-sided arrows axes of two pairs of jets, he suggested to be the main-jet axis of SNR~G0.9+0.1, and by two-sided double-lined red arrows, two additional symmetry axes he attributed to pairs of exploding jets. 
}
\label{fig:ThreeCCSNE}
\end{center}
\end{figure*}

In Figure \ref{Fig:CirC1} we present an image from \cite{Gasealahweetal2025}, with their marks in black. In our previous studies, we defined a bubble as a faint region fully or almost fully surrounded by a bright area, called a rim when it is narrow. We use the term ear to refer to a protrusion smaller than the main CCSNR shell, with a cross-section that decreases outward. As \cite{Gasealahweetal2025} noticed, the cross-section of the protrusion to the northwest (their NW bubble) increases somewhat, and only then decreases with distance from the center. We therefore use the term `fat ear.' We refer to their southeastern bubble as the SE ear. Following the morphology of the Cygnus Loop CCSNR, which we present in the middle panels of Figure \ref{fig:ThreeCCSNE}, we term the large protrusion to the south a blowout. In the upper panels of Figure \ref{fig:ThreeCCSNE}, we present two more images of the African Nebula from \cite{Gasealahweetal2025}. On panel (a) of Figure \ref{fig:ThreeCCSNE}, we identify four broken rims. A broken rim is characterized by a narrow, long, bright zone composed of two semi-straight filaments that run in different directions, almost touching each other. We suggest that each broken rim was shaped by a jet, which was directed along the narrow opening between the two semi-straight filaments. The two lower panels of Figure \ref{fig:ThreeCCSNE} show the CCSNR G0.9+0.1 with the rims in the large north ear (Ear N), and the blowout in the south. \cite{ShishkinKayeSoker2024} also identified a blowout in the MeerKAT radio emission map of the SNR G1.0-0.1 (from \citealt{HeywoodMeerKAT_G1.0-0.1_2022}); on the opposite side, there is an ear. 
           
  We identify the broken-rim structure by eye inspection, rather than a qualitative measure. Therefore, at this stage, it should be regarded as a speculative structure. More generally, to identify structural features, we follow the widely used practice of studying planetary nebula morphologies and classifying them. This is a careful eye inspection and qualitative classification (e.g., \citealt{Balick1987, Parkeretal2006, Sahaietal2007, Kwok2024}). This qualitative method enables the identification of jets' shaping (e.g., \citealt{SahaiTrauger1998}), comparison with numerical simulations, and sheds light on the shaping processes through qualitative comparisons with simulations (e.g., \citealt{GarciaSEguraetal2022, GarciaSEguraetal2025}) and by comparing planetary nebulae to CCSNRs (e.g., \citealt{Akashietal2018}). This method for identifying point-symmetric structures has been successfully applied to CCSNRs in the past (e.g., \citealt{Soker2024W44}).  

\begin{figure}[]
	\begin{center}
\includegraphics[trim=0.0cm 9.7cm 0.0cm 0.0cm ,clip, scale=0.46]{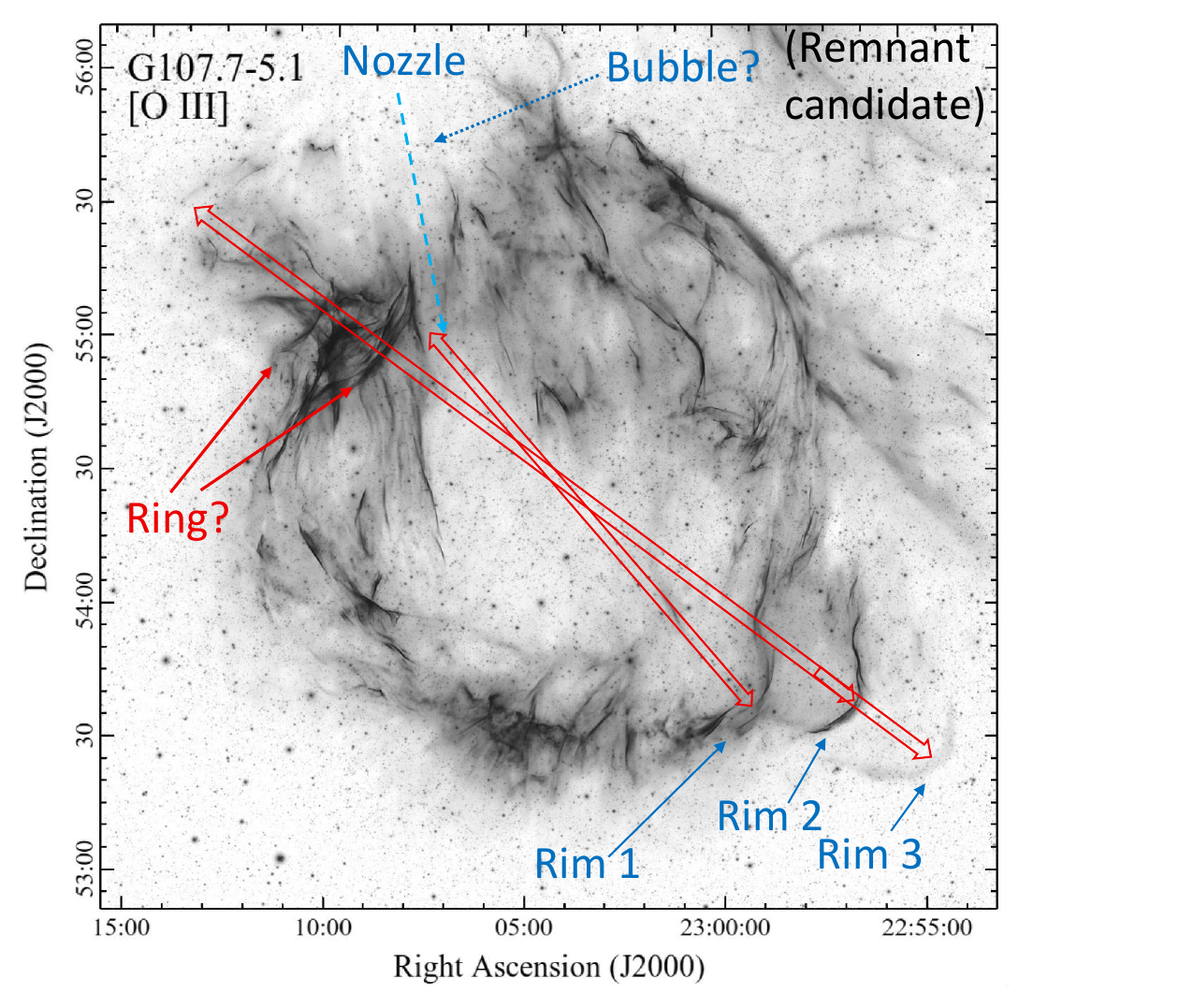} 
\caption{An image of a supernova remnant candidate G107.7-5.1 adapted from \cite{Fesenetal2024}, with the identified structural features of a bubble, a nozzle, and three rims (in blue) from \cite{Soker2024CF}. 
We identify a possible ring, through which we draw one symmetry axis shaped by two energetic pairs of jets. We suggest a second symmetry axis from Rim 1 to the nozzle.
}
\label{fig:G107}
\end{center}
\end{figure}

\cite{ShishkinKayeSoker2024} identified opposite structure features to the blowout of the Cygnus Loop: The north spout opposite to the south spout, and faint filaments that might be a north blowout (all marked on the middle panels of Figure \ref{fig:ThreeCCSNE}). In the CCSNR G0.9+0.1, there is a large ear in the north (Ear N), with its two rims, Rim 1 and Rim 2. Ear N is almost a blowout, but its symmetric shape makes it an ear rather than a blowout. In other astrophysical objects, jets inflate such rims inside ears and lobes, e.g., the three rims in the large ear of the planetary nebula KjPn 8 (\citealt{Lopezetal2000}; note that the other ear of KjPn 8 does not have such rims), and the four rims in one lobe of the galaxy Hercules A that appear in its radio image. Based on the similarity of the rims in one lobe of the CCSNR G107.7-5.1 (observed by \citealt{Fesenetal2024}) to the rims of KjPn 8 and Hercules A, \cite{Soker2024PNSN} argued that jets shaped the rims in the southwest ear of CCSNR G107.7-5.1; we present this CCSNR in Figure \ref{fig:G107} 

Based on these earlier arguments for jet inflating blowouts and shaping rims inside ears and lobes, we argue that three energetic jets shaped the blowout in the south of the Africa Nebula. Each jet shaped one broken rim. We further speculate that, at least in one case, there was an energetic enough counter (opposite) jet that shaped the broken rim in the north (marked BR N in panel a of Figure \ref{fig:ThreeCCSNE}). 

\cite{Gasealahweetal2025} resolved and discussed the rings in the NW ear and SE ear, which they attributed to shaping by jets. \cite{Gasealahweetal2025} simulated post-explosion jets in two dimensions and obtained an ear; however, they did not obtain a fat ear as the NW ear is, namely, the cross-section of their ear did not expand, and they did not reproduce a bright ring. We accept that jets shaped these rings. \cite{Akashietal2025}, for example, demonstrate with three-dimensional hydrodynamical simulations the formation of dense and prominent circumjet rings in conditions appropriate for planetary nebulae. Earlier studies have discussed circumjet rings in CCSNRs and compared them with those in other astrophysical objects. \cite{SokerShishkinW49B} identified large rings in the supernova remnant W49B; these rings are about half the size of the remnant.  \cite{SokerShishkinW49B} argued that these are circumjet rings, although the jets are long gone, and compare these rings to the rings of SNR 0540-69.3, rings that \cite{Soker2022SNR0540} argued to be circumjet rings. \cite{SokerShishkinW49B} compared the circumjet rings of these two CCSNRs to the rings around the active galactic nucleus jets in the galaxy Cygnus A, and the circumjet rings in the planetary nebulae MyCn 18 and Hen 2-104; all three objects exhibit jets. 

\begin{figure}[]
	\begin{center}
\includegraphics[trim=0.0cm 17.7cm 0.0cm 0.0cm ,clip, scale=0.98]{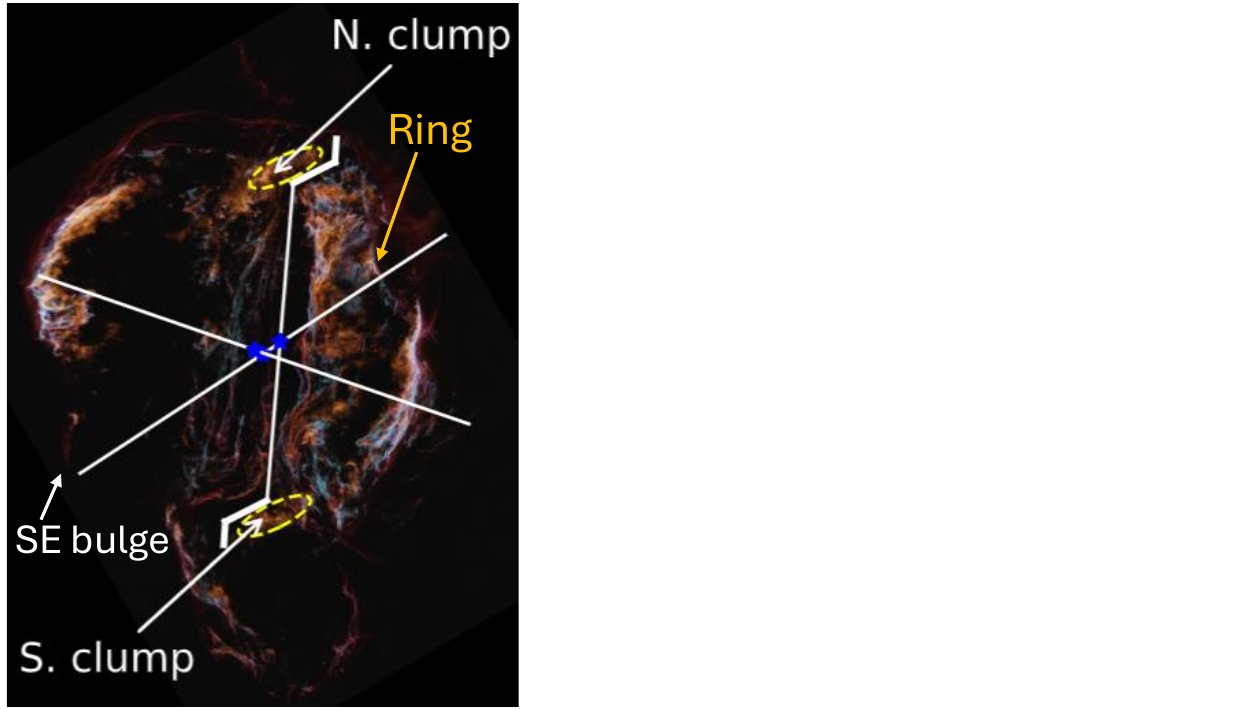} 
\caption{A visible-band image of the Cygnus Loop CCSNR adapted from \cite{Raymondetal2023} with marks in white from \cite{ShishkinKayeSoker2024}, and our specific identification of the ring (only part of which is observed), which they used to define the cavity. The three long white lines mark the claim by \cite{ShishkinKayeSoker2024} for three pairs of energetic jets that participated in the explosion of the Cygnus Loop. 
}
\label{fig:cygnus}
\end{center}
\end{figure}

Following the convincing evidence for circumjet rings in the Africa Nebula \citep{Gasealahweetal2025}, and its similarity to the Cygnus Loop, and SNR G107.7-5.1 that we pointed out above, we reexamined the last two for circumjet rings (although in these CCSNRs the jets are long gone). We note that \cite{ShishkinKayeSoker2024} identified the `cavity' in the Cygnus Loop based on a partial ring that is prominent in a Cygnus Loop image from \cite{Raymondetal2023}, which we present in Figure \ref{fig:cygnus}. 
The main point we make here is that while Cir X-1 is actively launching pairs of precessing jets inside the Africa Nebula, there are no such jets in the Cygnus Loop. As we claimed before \citep{ShishkinKayeSoker2024}, a jet that was part of the explosion process of the Cygnus Loop shaped the ring (cavity) in the Cygnus Loop. Other morphological features suggest a total of three energetic pairs of jets in the explosion process, as the white long lines indicate in Figure \ref{fig:cygnus} adapted from \cite{ShishkinKayeSoker2024}. Other, much weaker, pairs of jets have also likely contributed to the explosion.    

\cite{Soker2024CF} noted that the rim-nozzle asymmetry of CCSNR G107.7-5.1, presented in Figure \ref{fig:G107}, is similar to the rim-nozzle asymmetry in the cooling flow group of galaxies NGC 5813, which is known to be shaped by active galactic nucleus jets, and to some jet-shaped planetary nebulae. CCSNR G107.7-5.1 has one side of a large ear with two rims, and the other side has a nozzle. Recent studies emphasizing the importance of circumstellar rings in CCSNe (e.g., \citealt{SokerShishkinW49B, Gasealahweetal2025}) motivated us to reexamine CCSNR G107.7-5.1. As we mark in Figure \ref{fig:G107}, we now identify a possible ring in the northeast. 
With this identification, we suggest two symmetry axes to characterize CCSNR G107.7-5.1, as we draw by the double-lines double-sided red arrows in Figure \ref{fig:G107}. 
One axis connects Rim 1 to the nozzle. We suggest a pair of jets along this axis. The second axis running from Rim 3 through the ring and to the protrusion in the northeast was shaped by two pairs of energetic jets. We therefore solidify the identification of CCSNR G107.7-5.1 as a point-symmetric CCSNR. 
All these jets are from the explosion process, because the large ear with its two rims suggests very energetic jets, not just weak jets from a post-explosion low-mass fallback.

\section{The exploding jets of CCSNR Cir X-1}
\label{sec:jets}

We tentatively identify two main axes in the Africa Nebula that we attribute to at least four pairs of jets. 
We accept that post-explosion jets shaped, and still shape, the internal region around Cir X-1: the NW shock, SE shock, NE pocket, SW pocket, and the very inner jet pairs (e.g., \citealt{Tudoseetal2006, Selletal2010, Heinzetal2013, Coriatetal2019, Gasealahweetal2025}). 
 The distinction between explosion jets and binary post-explosion jets is in the time they are launched. Explosion jets are active during the explosion process, namely, as the stellar envelope is accelerated to velocities exceeding the escape velocity. Binary-driven jets are formed from binary interaction, which takes place after the explosion. In the case of Cir X-1, the energy that the binary jets carry is much lower than the explosion energy, and they shape regions deep inside the outer boundary of the ejecta. There are no clear structural features that link the inner structures of the pockets and shocks to the structures of the rings. Therefore, we take this to imply that there is no continuous activity from the explosion jets to the binary jets.     

While \cite{Gasealahweetal2025} attributed the NW ear and SE ear (their bubbles) to jets that the Cir X-1 binary systems launched years after the explosion, we attribute them to a pair of jets at the end of the explosion process. Our arguments are as follows. 

(1) \textit{Similarity to the cavity-ring in Cygnus Loop.} 
The Cygnus Loop has a ring on its northwest, which defines the cavity \citep{ShishkinKayeSoker2024}, and is similar to the NW ring of the Africa Nebula. No indications of any central jet source in the Cygnus Loop exist. The pulsar of SNR G0.9+0.1 launches jets as the insets on the lower panels of Figure \ref{fig:ThreeCCSNE} show. However, there are no indications that they shape the ears on the outskirts of this CCSNR and along the same direction. The pulsar jets are weak and bend in the inner region rather than shaping the ears. These similarities do not demonstrate that the explosion shaped the rings of the Africa Nebula, but they do raise this possibility. 

(2) \textit{The high energy of the post-explosion jets.} \cite{Gasealahweetal2025} find in their two-dimensional hydrodynamical simulations that the energy of the jet that shaped the NW ring, $1.1 \times 10^{50} \erg$, is about a third of the explosion energy in their simulations, $3 \times 10^{50} \erg$. This is a very energetic and powerful jet, with a power approximately 35 times the Eddington luminosity, according to simulations by \cite{Gasealahweetal2025}. 
Consider a similar energy to the counterjet; we obtain an extremely powerful post-explosion pair of jets, which is unlikely in X-ray binaries. These jets have the energy of energetic jet pairs in the JJEM (e.g., \citealt{Soker2025Learning}). 
We note that the jet that shaped the SE ear might also have been very energetic. The SE ear is small; however, a plume from this ear extends over a large distance. Such plumes that are the outcome of jets becoming subsonic and abruptly changing direction exit in planetary nebulae, e.g., the northern jet of the planetary nebula NGC 40 (e.g., \citealt{RodriguezGonzalezetal2022}). Such a plume exists in the CCSNR G309.2–00.6; \cite{ShishkinKayeSoker2024} termed the large volume that this plume covers a blowout. In light of these arguments, we suggest that the plume was formed by the jet that inflated the SE ear.    

(3) \textit{Condition for ring formation.} \cite{Akashietal2025} find that when a jet penetrates a dense shell, it compresses the shell to form a circumjet ring. The propagation of a strong explosion shock, as in CCSNe, through a steeply declining density profile, as exists in the exploding cores and envelopes of CCSNe, leaves a dense shell behind the shock (e.g., \citealt{Chevalier1976} for a general self-similar solution, and, e.g., \citealt{Onoetal200, Sandovaletal2021, ChenchenOno2024} for simulations that form a dense shell). 
We propose that at this late phase of the explosion, the newly born NS of Cir X-1 launched a pair of energetic jets that propagate through such a shell. The jets shaped the rings. The ears (bubbles) may be remnants of this interaction, or hot gas from the CCSNR that expanded through the rings at later times. These processes are the subject of ongoing research; we present preliminary results from that study in Section \ref{sec:Simulations}. 

None of these arguments precludes the post-explosion jets model of \cite{Gasealahweetal2025}, in which the Cir X-1 binary system launched relativistic jets years after the explosion that formed the ears (which they termed bubbles). Still, the arguments above collectively suggest a plausible alternative within the JJEM framework: that energetic jets at the final phase of the explosion process shaped the NW and SE rings of the Africa Nebula. We mark our suggested pair of jets that shaped the rings by the orange double-line arrows in panel (b) of Figure \ref{fig:ThreeCCSNE}.

We also identify a general axis that connects the north broken rim (BR N) to the blowout. In this case, we consider that three pairs of jets, only slightly inclined to one another, were highly asymmetric, with the southern jets much more powerful than the northern jets. Only one north jet left a signature in the remnant, the north broken rim. We mark our suggested energetic jets by the pale-blue double-lined arrows in panel (b) of Figure \ref{fig:ThreeCCSNE}.  

Overall, we argue that the morphology of the Africa Nebula (Cir X-1 SNR) hints at shaping by at least two energetic pairs of jets during the explosion process. The very late jets that shaped the rings were about equal in power. In the pairs of jets along the south-north axis, the southern jets, which we identify as at least three, were much more energetic than the northern jets.  
We, therefore, add the Africa Nebula to the group of point-symmetric CCSNRs. This group provides the strongest observational support for the JJEM and poses a significant challenge to the neutrino-driven explosion mechanism.   

\section{Ring formation at the explosion}
\label{sec:Simulations}

We present preliminary results from an extensive ongoing study of ring formation during the explosion process (Akashi and Soker, 2026, in preparation). 
In this preliminary study on the feasibility of jet-shell interaction during explosion-induced ring formation, we assume that the early phase of the explosion, namely, the earlier pairs of jittering jets, has already exploded the inner core and formed a dense, expanding shell. For simplicity and to present the basic ring formation process, we take a spherical expanding shell, but note that the same mechanism can work for a partial shell that is the front of a bubble, as earlier jets do inflate (see the three-dimensional JJEM simulations by \citealt{Braudoetal2025}). 

We perform three-dimensional hydrodynamical simulations with the Eulerian, adaptive-mesh refinement (AMR) code \textsc{FLASH} v4.8 \citep{FryxellEtAl2000}, using the unsplit hydrodynamics solver. The computational domain is a Cartesian box $(x,y,z)$ of size 
$L_x = L_y = L_z = 3.0 \times 10^{10} \cm$. From the center to each face of the cubical grid, the distance is $0.22 R_\odot$. We apply outflow boundary conditions on all faces of the cubical grid. We simulate two counterpropagating (opposite) jets launched with their axes aligned with the $z$-axis ($x=y=0$). The two jets may differ in power or half-opening angle. 

We use seven refinement levels above the base grid, and the jets' injection region is refined by one additional level (a total of eight AMR levels). This maximum refinement corresponds to an effective linear resolution of $2^{10}$ cells across the box and yields a minimum cell size in the jets' injection regions of 
$\Delta x_{\min} = {L}/{2^{10}} =293 \km$. We do not resolve the actual launching site of the jets, which is an order of magnitude smaller. We launch jets at larger radii, as is commonly done in similar simulations of other astrophysical systems, such as heating the intracluster gas in cooling-flow clusters and shaping planetary nebulae. 


The initial conditions, when we set $t=0$, include a dense and very fast-expanding spherical shell inside the core, as we assume that earlier jets already exploded the inner core of the stellar model; other calculations and simulations (not in the framework of the JJEM) of CCSNe obtained such a shell (e.g., \citealt{Chevalier1976, Onoetal200, Sandovaletal2021, ChenchenOno2024}). This shell has mass of $M_s = 4.5\,M_\odot$ and a radial outward velocity of $v_s = 5{,}000\ \mathrm{km\,s^{-1}}$, and it resides at $t=0$ inside $2.0 \times 10^{9} \cm < r < 3.0 \times 10^{9} \cm$. The initial kinetic energy of this shell is $E_{s,\mathrm{k}} = 1.12 \times 10^{51} \erg$. We do not include a full equation of state, but rather set an adiabatic index of $\gamma=4/3$, as appropriate for these conditions (e.g., \citealt{Chevalier1976}), with a uniform initial temperature of $T_s=10^8 \K$, giving an initial shell thermal energy of $E_{s,\mathrm{th}} = 3.6 \times 10^{50} \erg$.  
Outside this compressed shell, we take the density profile of the pre-explosion core from  \citep{Braudoetal2025}. This hydrogen- and helium-depleted progenitor has a radius of $R_\ast = 80,000 \km = 0.115 R_\odot$, which is well within our numerical grid.
To avoid numerical difficulties near the center of the grid, which is the center of the star, we build a numerical core at the center inside a radius of $r_{\rm core}=3 \times 10^8 \cm$, having a constant density and temperature ($\rho_{\rm core} = 10^{6}\ \mathrm{g}\mathrm{cm}^{-3}; 
T_{\rm core} = 2 \times 10^{8}\ \mathrm{K}$), which are not significant to this study because we do not model the explosion near the center. 
This inner core is purely numerical to prevent unrealistically small cell sizes and prohibitively small time steps. We exclude gravity in these simulations, reserving it for future full-scale simulations that will require significantly more time.  

We inject the jets from two opposite cones along the $+z$ and $-z$ directions, centered on the origin and inside a radius of $r_{\rm inj} = 9 \times 10^{8}\ \mathrm{cm}$. Their initial temperature is $T_j = 2 \times 10^{8}\ \mathrm{K}$, and their radial outward velocity is $v_j = 5 \times 10^{4} \km \s^{-1}$. In this study, all jets have the same activity period of $\Delta t_{\rm jet} = 0.05\ \mathrm{s}$. 
Because the jets are highly supersonic, their precise initial temperature has only a small effect on the large-scale dynamics; it mainly affects the immediate post-injection thermodynamics before adiabatic expansion.
We vary the energy and/or the half-opening angle of the jets, such that jets in a pair are not identical to each other. 

In this study, we do not analyze interactions or scan a large parameter space, as our goal is only to demonstrate the feasibility of late-explosion jets forming rings. To mimic observations, we present a map of the scaled emission integral of the numerical results, which is the integration of the density square along the line of sight: $\int \rho^{2} d Y_s$, where $Y_s$ is the coordinate along the line of sight. We remove the inner core from the calculation of the scaled emission integral as it is not physically meaningful, and we do not include late jittering jets or gravity, both of which play a role in the inner regions. We are interested only in the expanding outer ejecta.    

\begin{figure}[]
	\begin{center}
\includegraphics[trim=4.0cm 1.4cm 4.2cm 0.0cm ,clip, scale=0.70]{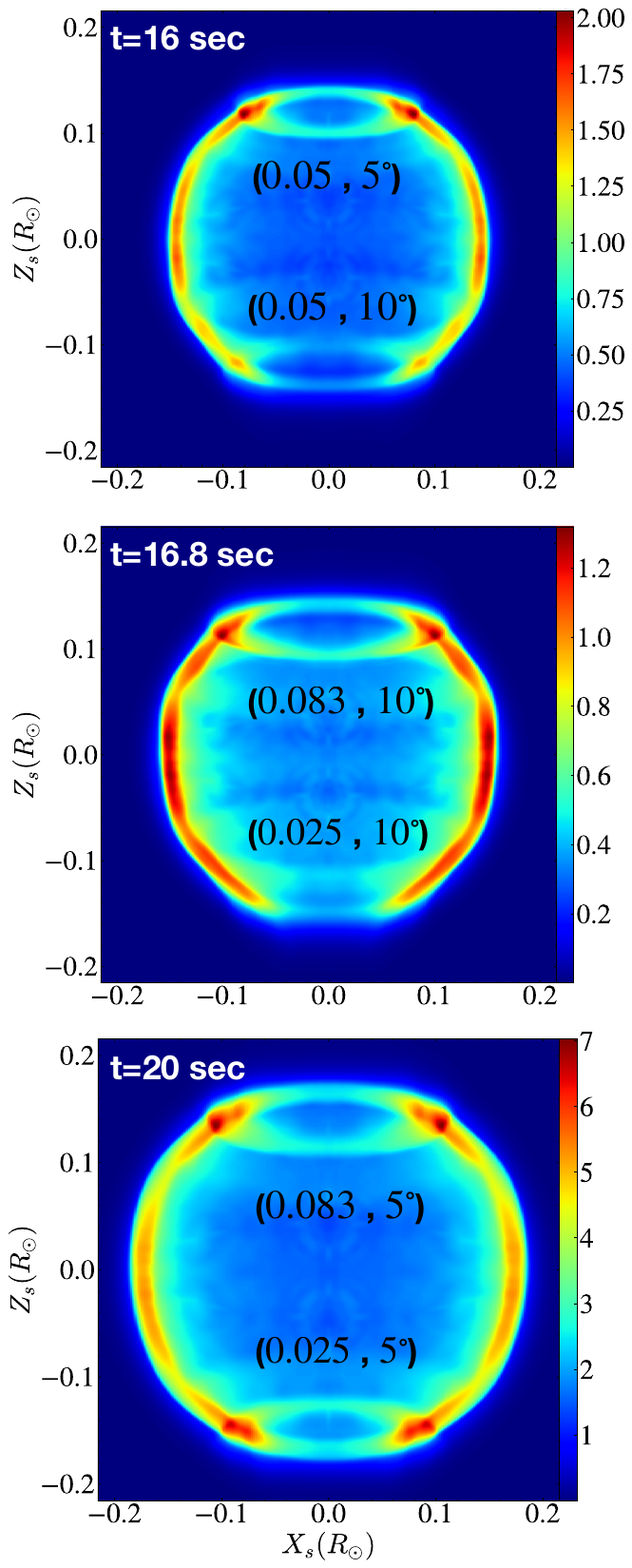} 
\caption{Scaled emission integral maps of three simulations. The scaled emission integral is the integral of the square of the density along the line of sight, $\int \rho^2 d Y_s$, and it mimics observations. Each simulation shows the formation of two opposing rings produced by two unequal jets at the final phase of the explosion. In each panel, we record the time at which the simulation was terminated, the energy of each jet in units of $10^{51} \erg$, and its half-opening angle in degrees. In all cases, the axis of the two jets is inclined at $75^\circ$ to the line of sight. The coordinates on the plane of the sky are $X_s$, which coincides with the grid's $x$-axis, and $Z_s$, which is in the grid's $yz$ plane and at $15^\circ$ to the $z$ axis (the jets' axis). 
The color bar of the upper two panels is in units of a  $10^{17}\ \mathrm{g}^2\mathrm{cm}^{-5}$, while the color bar of the third panel is in units of $10^{16}\ \mathrm{g}^2\mathrm{cm}^{-5}$.
}
\label{fig:rings_Africa}
\end{center}
\end{figure}

In Figure \ref{fig:rings_Africa}, we present the scaled emission integral maps of three simulations, i.e., the integration over the density squared along the line of sight. These panels mimic the observations. In all cases, the jets' axis is tilted at $75^\circ$ degrees to the line of sight (the jets' axis is $15^\circ$ to the plane of the sky). The coordinates are those on the plane of the sky: $X_s$ coincides with the grid's $x$-axis, and $Z_s$ is in the $x=0$ plane and tilted by $15^\circ$ to the grid's $z$-axis. For each panel, we record the time at which the simulation was terminated, either when the flow reached the grid edges or when numerical difficulties were encountered. For each jet, we give the energy in units of $10^{51} \erg$ and the initial half-opening angle in degrees: $(E_j/10^{51} \erg, \alpha_j)$. The three panels of Figure \ref{fig:rings_Africa} present clear circumjet rings.
The typical ratio of the kinetic energy to thermal energy of the ejecta at the end of the simulations is in the range of $8 - 10$. Since the kinetic energy is much larger than the thermal energy at the end of our simulations, the dense shell and rings will maintain their morphology as they expand further.  

Our conclusion from the simple, preliminary three-dimensional simulations is that jets in the final phase of the explosion process can form rings within the expanding shell. There is still much to do (Akashi and Soker 2026, in preparation), including a more accurate equation of state, that includes radiation pressure and gas pressure, the self-consistent explosion of the core with jittering jets, including gravity, continuing the simulations to much later phases when the ejecta interacts with the circumstellar material and a reverse shock heats the interior, and including the activity of the Cir X-1 binary system.  These simulations might lead to a better reproduction of the observations, particularly the ear, e.g., the faint protrusion from the NW ring (the NW bubble in Figure \ref{Fig:CirC1}), which we could not reproduce with the present simple simulations. 

We aimed to assess the feasibility of jet-induced ring formation during the explosion process, and we successfully achieved this. With the present simple set of simulations, however, we did not produce the ear, i.e., the protrusion through the ring. In future simulations, we will include additional effects that we expect to contribute to ear formation. These ingredients are: turning off the jets slowly so that the last, weak jets may form the ear; interaction with circumstellar material that confines the protruding jets, thereby forming the ear; and a later formation of the ear. The last process involves a high-pressure gas from the CCSNR streaming out from the ring at late times to form the ear. In turn, the high-pressure gas inside the CCSNR results from the reverse shock that heats the interior, and, in cases where there is an active NS, as in the Cir X-1 binary system, from the deposition of energy in the inner regions by the NS. We speculate that this final process drives CCSNR material from the interior through the ring.

\section{The kick velocity of Cir X-1}
\label{sec:kick}

In this section, we discuss the kick-BEAP, namely, the kick velocity imparted to the NS by pairs of unequal jets at the explosion. We show that the parameters of the Cir X-1 binary system imply that the kick-BEAP can operate without destroying the binary system. We consider a plausible set of parameters for the pre-explosion binary system, but admit it is not unique. 
 
The large blowout suggests that the jets that inflated it, in the frame of the JJEM, were very energetic, tens of percent of the explosion energy,  $E_{\rm BOj}>10^{50} \erg$. 
The jets that inflated the blowout were much more energetic than the opposite jets and carried much higher momentum. 
The NS launched these jets via intermittent accretion disks, and, by momentum conservation, the jets imparted a kick velocity to the NS in the opposite direction; this is the kick-BEAP mechanism. The NS kick velocity by the kick-BEAP mechanism is  \begin{equation}
\begin{split}
\Delta v_{\rm NS,kP} & = 134
\left( \frac{E_{\rm BOj}}{ 2 \times 10^{50} \erg} \right)  
\\ & \times 
\left( \frac{v_{\rm j}}{10^5 \km \s^{-1}} \right) ^{-1}
\left( \frac{M_{\rm NS}}{1.5 M_\odot} \right)^{-1}
\km \s^{-1}, 
\label{eq:V1jet}
\end{split}
\end{equation}

The Cir X-1 is an unusual X-ray binary system in having a more massive companion to the NS than in typical low-mass X-ray binaries and in having a high eccentricity (e.g., \citealt{Iariaetal2005}), although the values are uncertain. \cite{Johnstonetal1999} take $M_2 \simeq 3-5 M_\odot$ for the companion mass and $e \approx 0.7 - 0.9$ for the eccentricity, while \cite{Rankinetal2024} quote a value of $e \approx 0.45$. 
We will take the upper mass range for the companion, $M_2=5 M_\odot$, to account for the slow proper motion of the Cir X-1 binary system. We also scale by an NS mass of $M_{\rm NS}=1.5 M_\odot$.   
There is an agreement among the different studies on the orbital period of $P \simeq 16.5$~day.  
For this orbital period, the present semi-major axis of the system is $a_b = 51 (M/6.5M_\odot)^{1/3} R_\odot$. The periastron for $e \simeq 0.5$ is $\simeq 25 R_\odot$. If we assume this distance to be the pre-explosion semi-major axis and a circular orbit, the relative pre-explosion orbital velocity was $v_0 = 246 (M_0/8M_\odot)^{1/2} (a_0/25R_\odot)^{-1/2} \km \s^{-1}$, where $M_0$ is the pre-explosion binary system mass.   
For these parameters, the mass of the NS progenitor was $3 M_\odot$ at explosion, and the ejected mass was $M_{\rm ej} \simeq 1.5 M_\odot$, namely, the CCSN was a type Ic CCSN with low ejected mass. 
   

For the above parameters, the pre-explosion orbital velocity of the NS progenitor, i.e., the exploding star, around the center of mass was $v_{\rm pe,NS} \simeq 154 \km \s^{-1}$. The ejecta leaves the system at the NS progenitor velocity. 
From momentum conservation, the ejection of this mass imparted a kick to the binary system in the opposite direction and with a magnitude of $v_{\rm k-ej} = (1.5/6.5) v_{\rm pe,NS} \simeq 35 \km \s^{-1}$. 

To avoid unbinding the system, the kick imparted by the jets, the kick-BEAP, cannot be aligned with the pre-explosion velocity of the NS (otherwise, the relative velocity of the NS and the companion is too large to maintain the system bound). It should have a large component to counter the NS velocity. Namely, the kick-BEAP imparted a kick velocity to the binary system, more or less, in the same direction as that due to the mass ejection in the explosion. With the parameters of the kick-BEAP by equation (\ref{eq:V1jet}), the kick-BEAP imparted to the binary system a kick velocity of  $v_{\rm k-P} \simeq (1.5/6.5) \Delta v_{\rm NS,kP}  \simeq 31 \km \s^{-1}$. The two kick velocities might add up to a value of
\begin{equation}
v_{\rm B} \simeq v_{\rm k-e} + v_{\rm k-P} < 66 \km \s^{-1}. 
\label{eq:BinaryKick}
\end{equation}
Since they will not be exactly in the same direction, the vector addition of these two kick velocities will be smaller than the sum of the two values.   
This is the total kick velocity of the binary system. The proper velocity is smaller than this value if the kick is not on the plane of the sky. 

The proper motion of the Cir X-1 binary system is highly uncertain. 
\cite{Mignanietal2002} derive a proper motion of $1.5 \pm 1.6~{\rm mas} \yr^{-1}$. This is compatible with a zero proper motion, or, in any case, a low value of $v_{\rm Cir}\simeq 65 \km \s^{-1}$ for a distance of $D_{\rm CirX-1} \simeq 9.4 \kpc$;
for the distance we take $D_{\rm CirX-1} \simeq 9.4 \kpc$ from \citealt{Heinzetal2015}, like \cite{Gasealahweetal2025} do (lower values exist in earlier studies). 
The results of \cite{Tudoseetal2008} are consistent with slow proper motion, but they give only a large upper limit. 

We conclude that the kick-BEAP and the kick by mass ejection at the explosion, as we derived in equation (\ref{eq:BinaryKick}), are compatible with the limit on the proper motion of the Cir X-1 binary system. 
Again, the above set of parameters is only a plausible set; we do not claim that it must be the set of parameters of the system. A more detailed study will be conducted after numerical simulations determine the energy of the jets required to inflate the blowout of the Africa Nebula within the framework of the JJEM.  

In the scenario we considered in the simulations of Section \ref{sec:Simulations} and above, the Africa Nebula was a striped-envelope CCSN, a Type Ic CCSN. This deserves further study. We note that close binary systems can form hydrogen and helium-stripped CCSN progenitors (e.g., \citealt{Gilkisetal2025}).

\section{Summary} 
\label{sec:Summary}

We studied and discussed a scenario according to which a pair of opposite jets at a very late phase of the explosion process shaped the NW and SE rings of the Cir X-1 CCSNR, termed the Africa Nebula by \cite{Gasealahweetal2025}, who found the ears and resolved the rings (Figure \ref{Fig:CirC1}). The Cir X-1 binary system is actively launching jets, possibly precessing jets that are shaping the inner point-symmetric structure of the Africa Nebulae near the binary system. \cite{Gasealahweetal2025} proposed that jets that the binary system launched years after the explosion, shaped the rings and the ears. 

We proposed that the newly born NS launched a pair of very late jets, which are still part of the explosion process, namely, several seconds after the earlier jets in the frame of the JJEM had set up a shock in the core. This shock compressed an expanding dense shell. The two very late jets penetrated this shell, a process that can shape circumjet rings and drive material farther out (e.g., \citealt{Akashietal2025}). We base this scenario on three arguments (Section \ref{sec:jets}): (1) Similarity to other CCSNe that have no presently active jets that can shape ears and rings on the outskirts of the CCSNR shell (Figure \ref{fig:ThreeCCSNE}), particularly the Cygnus Loop (Figure \ref{fig:cygnus}). (2)  
\cite{Gasealahweetal2025} found that their proposed post-explosion jets need to be extremely energetic for an X-ray binary system; this energy is typical for JJEM jets. (3) The above-mentioned process of jets penetrating a dense shell, and that a dense shell develops in the core shortly after the explosion has started. In Section \ref{sec:Simulations} we described preliminary three-dimensional hydrodynamical simulations that show the feasibility of this scenario to shape rings during the explosion process. The three simulations we present in Figure \ref{fig:rings_Africa} present clear rings in the expanding shell. Our simulations did not produce the ears protruding from the rings. We speculate that the ears are later expanding gas. As the reverse shock and the activity of the Cir X-1 binary system heat the inner ejecta of the CCSNR, the pressure rises, and the ejecta expand outward from the ring. This is a subject of a future study.  At this point, the formation of the rings by late (post-explosion) jets as \cite{Gasealahweetal2025} proposed, should also be considered a possible explanation for the rings.

We identified four broken rims, i.e., composed of two touching semi-straight filaments, in the Africa Nebula; two in the main shell and two in the blowout (panel a of Figure \ref{fig:ThreeCCSNE}). From the similarity of the blowout of the Africa Nebula to that of the Cygnus Loop (Figure \ref{fig:ThreeCCSNE}), and the similarity of the rims inside the blowout of the Africa Nebula to the rims in the large ears of SNR G0.9+0.1 (Figure \ref{fig:ThreeCCSNE}) and SNR G107.7-5.1 (Figure \ref{fig:G107}), we  speculated that exploding jets, namely, jets that participated in the explosion process in the framework of the JJEM, shaped the blowout and the broken rims. 
We mark the six energetic exploding jets that we proposed by the double-lined arrows in panel b of Figure \ref{fig:ThreeCCSNE}. The JJEM allows for more exploding jets, but they left no marks on the remnant, because they were probably weaker. 

Based on the enlightening discussion by \cite{Gasealahweetal2025} of the rings in the Africa Nebula, we revisited the CCSNR G107.7-5.1. We tentatively identified a ring there and argued that the morphology of this CCSNR reveals two symmetry axes, as the double-lined double-sided arrows show in Figure \ref{fig:G107}. This strengthens the claim that SNR G107.7-5.1 is a point-symmetric CCSNR. 

This study, above all, further demonstrated the usage of the JJEM in analyzing CCSNRs. The JJEM is a successful explosion mechanism for analyzing CCSNRs and CCSNe (see the references in Section \ref{sec:intro}). For example, the JJEM can naturally explain the double-peak emission from iron and intermediate elements in the CCSN observed by \cite{Hueichapanetal2025}; according to the JJEM, pairs of opposite jets can account for this observation.  The JJEM can also account for multiple shells in the photospheric phase of CCSNe \citep{SokerShiran2025, ShiranSoker2026}. That said, there are still challenges that simulations of the JJEM should address, e.g., improved matches with observed rings and ears, and the formation of intermittent accretion disks required to launch jets.

\section*{Acknowledgements}

We thank Dima Shishkin for helpful discussion and suggestions,  and an anonymous referee for useful comments.  
NS thanks the Charles Wolfson Academic Chair at the Technion for the support.





 \bibliography{reference}{}
  \bibliographystyle{aasjournal}

\end{document}